\documentclass[journal=jpccck,manuscript=article]{achemso}
\usepackage[version=3]{mhchem} 

\usepackage{graphicx}
\usepackage{amsmath}
\usepackage{color}
\usepackage{multirow}
\newcommand{\onlinecite}[1]{\hspace{-1 ex} \nocite{#1}\citenum{#1}}

\author{Linggang Zhu}
\affiliation{SUPA, School of Physics and Astronomy, The University of Edinburgh, Edinburgh EH9 3JZ, UK}
\affiliation{Shenyang National Laboratory for Materials Science, Institute of Metal Research, Chinese Academy of Sciences, Shenyang 110016, China}

\author{Qing-Miao Hu}
\author{Rui Yang}
\affiliation{Shenyang National Laboratory for Materials Science, Institute of Metal Research, Chinese Academy of Sciences, Shenyang 110016, China}

\author{Graeme~J.~Ackland}
\affiliation{SUPA, School of Physics and Astronomy, The University of Edinburgh, Edinburgh EH9 3JZ, UK}
\email{gjackland@ed.ac.uk}

\title{Binding of an oxide layer to a metal: the case of
  Ti(10$\overline{1}$0)/TiO$_{2}$(100)}

\begin{document}
\begin{abstract}
We study the chemical nature of the bonding of an oxide layer to the
parent metal. In order to disentangle chemical effects from
strain/misfit, Ti(10$\overline{1}$0)/TiO$_{2}$(100) interface has been
chosen.  We use the density functional pseudopotential method which
gives good agreement with experiment for known properties of bulk and
surface Ti and TiO$_2$. Two geometries, a film-like model (with free
surface in the structure) and a bulk-like model (with no free surface
in the structure) are used to simulate the interface, in each case
with different terminations of Ti and TiO$_2$. For the single-oxygen interfaces, the interface energy obtained using these two models agree
with each other; however for the double-oxygen ones, the relative
stability is quite different.  The disturbance to the electronic
structure is confined within a few atomic layers of the interface.
The interfacial bonding is mainly ionic, and surprisingly there is
more charge transfer from Ti to O in the interface than in the bulk.
In consequence the Ti/TiO$_2$ interface has stronger binding than the
bulk of either material.  This helps to explain why the oxide
forms a stable, protective layer on Ti and Ti alloys.
\end{abstract}

\section{INTRODUCTION}

In contrast to the many atomistic studies of metals and oxides, both
at the ab initio and empirical level, the metal-oxide interface is
much less well studied.  This may appear surprising given its central
role in corrosion.  One reason may be the difficulty in identifying a
candidate interface where the chemical effects are not entangled with
misfit strain energy.  Another is the relatively large size of system
which is required to isolate the system.  Although the strain energy
is long-ranged, there are reasons to expect that the chemical effects
of the interface may not be: on the metal side the free electrons
should screen the Coulomb forces, while the image charges induced in
the metal should mimic the electrostatics of a bulk oxide.  At the
interface itself, such classical ideas break down, and a full quantum
treatment is needed to determine the nature of the bonding.  In this
paper we will consider the low-misfit
Ti(10$\overline{1}$0)/TiO$_{2}$(100) interface, with a view to
determining the range over which chemical effects are significant and
the nature of the bonding cross the interface.

Titanium alloys are widely used in many fields such as the aerospace industry, chemical plants, and even sporting goods. The high strength-to-weight ratio and excellent corrosion resistance account for its wide application. The corrosion property is mainly a result of the formation of stable protective oxide film, which consists primarily of TiO$_{2}$. However, above 600$^\circ$C, the fast diffusion of oxygen through the oxide layer into the bulk can result in excessive growth of oxide layer and embrittlement of the adjacent oxygen rich layer of the titanium alloy, which limits its maximum use temperature.\cite{Titanium} Alloying and coating have been found effective to address this problem. Obviously, the interface between the titanium and its oxide plays a vital role in the corrosion, both through its adhesive strength and the diffusion of species (O and/or Ti) through it, and its structure is a key aspect for understanding the behavior of titanium alloys.

Extensive experiments on pure titanium\cite{Flower1974,Kozlowski1988,Ting2002,Kumar2010} and titanium alloys\cite{Lopez2003, Garcia2003} have established that the crystal structure of the oxide is   normally rutile (tetragonal, P4$_{2}$/mnm). Although Guleryuz et al.\cite{Guleryuz2004} reported some diffraction angles consistent with the  anatase structure(tetragonal, I4$_{1}$/amd), in the scale of Ti-6Al-4V oxidized at 600$^\circ$C, rutile is the dominant oxide at 650$^\circ$C. Due to the competition between surface free energy and strain energy, the growth of rutile on pure titanium exhibits a preferential direction,\cite{Kozlowski1988,Ting2002} with a specific crystallographic orientation relationship (COR) between titanium and rutile. Three possible CORs between Ti and rutile were established by Flower et al.\cite{Flower1974} using an \emph{in situ} method: Ti(0001)[11$\overline{2}$0] // TiO$_{2}$(010)[001], Ti(10$\overline{1}$0)[0001] // TiO$_{2}$(100)[010] and Ti(11$\overline{2}$0)[0001] // TiO$_{2}$(001)[100]. Among these three different matchings, Ti(10$\overline{1}$0)[0001] // TiO$_{2}$(100)[010] has the smallest mismatch on the plane of the interface, which means only a tiny interface strain would be required.

The structure of the oxide and the orientation relationship can be easily determined by the experiments; however, some other quantities like the chemical composition, atomic structure of the interface, and the nature of the bonding (ionic/metallic/covalent) across the interface are currently experimentally inaccessible. Fortunately, nowadays, it is possible to deal with coherent interface structures (large system, low symmetry) using accurate first-principles theoretical methods to obtain those quantities. Coherency implies that one part (metal or oxide) will be strained to match the other one perfectly to maintain the coherency without misfit dislocations. In this case, DFT supercell method can be a good tool to study interfaces with small mismatch, but even when the misfit is quite big, it has been assumed that the interfacial regions between the misfit dislocation being modelled.\cite{Siegel2002} Such a first-principles supercell method has been used to study many different interfaces, for example, interface of ZrO$_{2}$/Ni,\cite{Christ2001} Al/Al$_{2}$O$_{3}$,\cite{Siegel2002} Ti/TiN,\cite{Liu2003} YSZ/Al$_{2}$O$_{3}$,\cite{Lallet2009} Nb/Nb$_{5}$Si$_{3}$,\cite{Shang2010} etc.  Here we present a DFT study of Ti(10$\overline{1}$0)[0001] // TiO$_{2}$(100)[010] interface. The purpose of this study is to determine the optimal atomic structure and energy of Ti(10$\overline{1}$0)[0001] // TiO$_{2}$(100)[010] interface, and characterize the nature of the interfacial bonding.

\section{METHODS}

The Vienna Ab initio Simulation Package (VASP)\cite{Kresse1993, Kresse1996, KresseG1996} utilizing a plane-wave basis set for the expansion of the single-particle Kohn-Sham wave functions, was used in this study. The projector augmented wave (PAW) method\cite{Blochl1994}, was employed to describe the electron-core interaction. The 3\emph{p} semicore electrons of Ti were treated as valence, given a 10-electron PAW-pseudopotential. For the exchange-correlation interaction, we adopted generalized gradient approximation (GGA) as parameterized by Perdew and Wang (PW91).\cite{Perdew1992} A high cutoff energy of 525 eV was used. Sampling of the Brillouin zone was performed with a Monkhorst-Pack grid.\cite{Monkhorst1976} Ground-state atomic structures were obtained by minimizing the Hellmann-Feynman forces on the atoms, and all the atoms were free to relax. The relaxations terminate when the maximum force on the atoms is less than 0.05 eV/\AA. For some calculations a dipole moment is present, in such cases the divergent terms are removed by the Ewald sum; we did not include a dipole correction\cite{Bengtsson1999} as previous work in TiO$_2$ has shown this to have minor effects.

\section{RESULTS: parameters and basic properties of Ti and TiO$_2$}

\subsection{Bulk properties}

\begin{table}
\caption{Structure parameters (internal coordinate u in TiO$_{2}$ refers to the position of the oxygen atom in the unit cell), bulk modulus $B_{0}$ (GPa), compared with experimental data.}
\label{tbl:1}
\begin{tabular}{l c c c | c c c c}
   & \multicolumn{3}{c}{HCP-Ti} & \multicolumn{4}{c}{TiO$_{2}$} \\
   & a(\AA) & c(\AA) & $B_{0}$ & a(\AA) & c(\AA)&  u  & $B_{0}$\\ \hline
 This work. & 2.934 & 4.638 & 115 & 4.645 & 2.971 & 0.305 & 206 \\
 Expt.\textsuperscript{\emph{a}}  & 2.951 & 4.674 & 110 & 4.587 & 2.954& 0.305 & 216\\
\end{tabular} \\
\leftline{\textsuperscript{\emph{a}} Data for Ti and TiO$_{2}$ are from refs \onlinecite{Kittle1971} and \onlinecite{Burdett1987}}
\end{table}

\begin{figure}
\includegraphics[width=7cm]{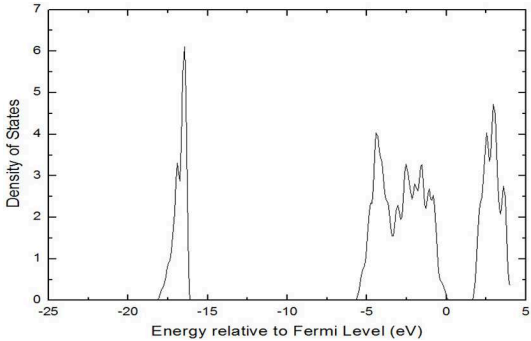}
\caption{DOS per TiO$_{2}$ unit in bulk rutile.\label{fig:1}}
\end{figure}

To verify the accuracy of our computational parameters, we first calculated the bulk properties of Ti and rutile. The k-point mesh was set at $5\times5\times8$ and  $11\times11\times7$ for the bulk TiO$_{2}$ and hcp-Ti, respectively. These provide convergence to within 1 meV and as can be seen from \ref{tbl:1} the calculated lattice parameter and bulk modulus data agree excellently with the experiments. From the DOS of TiO$_{2}$ (\ref{fig:1}), the gap between highest occupied and lowest unoccupied state of TiO$_2$ is about 1.7 eV, far away from the experimental value, which is around 3.0 eV\cite{Grant1959}: this does not affect the bonding and is typical of the error of DFT in describing excited states.

\subsection{Surface properties}

To determine the minimum slab thickness needed to reliably calculate the surface/interface we examined the convergence of the relaxation and energy of the surfaces with respect to slab thickness. In this calculation, we used k-point sampling $8\times5\times1$ and $5\times8\times1$ for Ti(10$\overline{1}$0) and TiO$_{2}$(100). In both slabs, a 15 \AA~vacuum region is introduced to avoid the interaction between periodic images. For both surfaces, only the (1 $\times$ 1) structures (along the two lattice vectors of the surface, the symmetry is the same as in the bulk) are considered, since these have been observed by for clean surface under normal conditions by experimentalists.\cite{Watson1989,Mischenko1989,Diebold2003}

The calculated surface energy might diverge with the thickness of the slab, if there is any numerical difference between the calculation of bulk and slab, arising, for example, from the k-point mesh. Following
Boettger\cite{Boettger1994} we avoid this possibility, by evaluating the surface energy $\gamma$  as follows:

\begin{equation}
\begin{split}
 \gamma = (E^{\rm{N}}_{\rm{slab}}-\rm{N}{\Delta} \emph{E}^{\rm{N'}}_{\rm{slab}}/\Delta{\rm{N}})/{2S} \\
 \Delta E^{\rm{N'}}_{\rm{slab}} = E^{\rm{N'}}_{\rm{slab}} - E^{\rm{N'-\Delta{N}}}_{\rm{slab}}
\end{split}
\end{equation}
where $\rm{N'}$ is the minimum number of the slab layers for which the energy converges, $E^{\rm{N}}_{\rm{slab}}$ is the energy of a N-layer slab, and S is the area of the surface.

\subsubsection{Ti(10\emph{$\overline{\emph{1}}$}0)}

In hcp Ti,  two different (10$\overline{1}$0) planes exist, depending on whether the surface terminates in a large (1.694 \AA) or small (0.847 \AA) interlayer spacing. From the experimental results, \cite{Mischenko1989} Ti(10$\overline{1}$0) surface with a small first interlayer spacing was the favored one. Our calculation confirmed this, and henceforth, we concentrate on the more stable Ti(10$\overline{1}$0) surface.

\begin{table}
\caption{Interlayer distance relaxation of the Ti(10$\overline{1}$0) surface vs slab thickness, shown as a percentage of the bulk spacing. The surface energy $\gamma$ obtained with different slab thickness is also listed, in J/m$^2$.}
\label{tbl:2}
\begin{tabular}{l  c c c c c c c c }
   Layer & $\Delta$$d_{12}$ & $\Delta$$d_{23}$  & $\Delta$$d_{34}$ & $\Delta$$d_{45}$ & $\Delta$$d_{56}$ & $\Delta$$d_{78}$ & $\Delta$$d_{89}$ & $\gamma$ \\ \hline
    12 & -3.26 & -5.43 & 5.98 & -2.20 & 3.65 & & & 1.99\\
    14 & -3.64 & -4.74 & 4.10 & -1.13 & 2.19 & 2.16 & & 1.99\\
    16 & -3.75 & -4.80 & 3.95 & -1.01 & 1.97 & 1.25 & -0.29 & 1.99 \\
    18 & -3.70 & -4.93 & 3.99 & -1.18 & 2.07 & 1.39 & -0.39  & 1.99\\
    Expt.\textsuperscript{\emph{a}} & -5.9 & & & & & & & \\
\end{tabular}\\
\leftline{\textsuperscript{\emph{a}} ref~\onlinecite{Mischenko1989}}
\end{table}

The surface relaxation data is shown in \ref{tbl:2}, the distance between the first two layers $d_{12}$ decreases by about 4\% compared with that in the bulk, in reasonable agreement with 6\% from the experiment \cite{Mischenko1989} (the discrepancy is less than 0.02 \AA). $d_{23}$ contracts and $d_{34}$ expands, showing the same trend as the results by self-consistent tight bonding model.\cite{Erdin2005} Relaxations between deeper layers are even smaller (less than 2\%). The surface energy  converged quickly with cell size to 1.99 J/m$^2$. This is slightly higher than the 1.91 J/m$^2$ of Ti(0001) surface which we obtained using the same method: Ti(0001) has an experimental energy of 1.99 J/m$^2$.\cite{Tyson1977} Thus Ti(10$\overline{1}$0) is just slightly less stable than the close-packed Ti(0001) surface. For further study, we chose a 16-layer slab to simulate Ti(10$\overline{1}$0).

\subsubsection{TiO$_{2}$(100)}

\begin{figure}
\includegraphics[width=5cm]{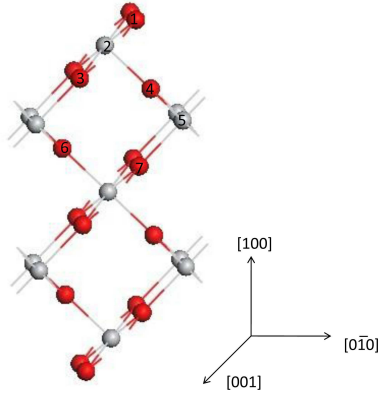}
\caption{TiO$_{2}$(100) surface structure. Red spheres represent O atoms, and silver spheres represent Ti atoms.\label{fig:2}}
\end{figure}

Extensive theoretical studies have been done to investigate the TiO$_{2}$(100) surfaces. There are three possible terminations for TiO$_{2}$(100), which we call O-Ti-O, O-O-Ti, and Ti-O-O (named by the atomic arrangement from the surface to inner layer). The first of these (\ref{fig:2}) has the smallest surface polarization due to the symmetrical arrangement of O, and the geometry has been observed by experiment.\cite{Diebold2003} We find it to be the most stable one, and we use the O-Ti-O terminated  TiO$_{2}$(100) surface to determine the minimum thickness of the slab needed. One thing that should be noted is that in this subsection, the O-Ti-O unit is treated as one layer, i.e., the 5-layer slab mentioned here has 15 atomic layers.

The choice of exchange-correlation functional has been found to affect the surface energy quite significantly,\cite{Muscat1999} and normally LDA gives a higher surface energy than GGA. Due to the lack of the experimental results, it is difficult to clarify which function best describes the real situation. Here, we compared our results with other calculations using the same exchange-correlation function (GGA), in order to verify the accuracy of our calculation. As shown in \ref{tbl:3}, O$_{1}$, Ti$_{2}$, and O$_{3}$ exhibit large relaxations along [0$\overline{1}$0]. Moreover, Ti$_{2}$ relaxes inward along [100], while outward relaxation is observed for O$_{1}$ and O$_{3}$. These relaxations increase the effective coordination of Ti$_{2}$ (fivefold).\cite{Muscat1999,Labat2008} Our surface relaxation results agree with that found by Muscat et al..\cite{Muscat1999} \ref{tbl:4} lists the surface energy dependence on the slab thickness: it converges quickly to 0.68 J/m$^2$ which agrees very well with Perron et al. \cite{Perron2007} (10 valence electrons are considered for Ti) and Labat et al. (GGA-PBE),\cite{Labat2008} while it is smaller than Muscat's GGA result\cite{Muscat1999}. Considering the convergence of both the surface structure and energy, we believe that a 13-layer slab is thick enough to model TiO$_{2}$(100).

\begin{table*}
\caption{The displacement of ions (\AA) (as labeled in \ref{fig:2}) relative to the bulk terminated structure of TiO$_{2}$(100) surface obtained with different slab thickness.}
\label{tbl:3}
\begin{tabular}{l c  c c c c  c c c c c }
  & \multicolumn{2}{c}{5}  & \multicolumn{2}{c}{9}   & \multicolumn{2}{c}{13}   & \multicolumn{2}{c}{17}   & \multicolumn{2}{c}{Ref.\textsuperscript{\emph{a}}}\\
  Label  & [0$\overline{1}$0] & [100] & [0$\overline{1}$0]  & [100] & [0$\overline{1}$0] & [100] & [0$\overline{1}$0] & [100] &  [0$\overline{1}$0] & [100] \\\hline
  1 & 0.32 & 0.07  & 0.32 & 0.08 &  0.28 & 0.06 & 0.28 & 0.06            &0.31&0.06\\
  2 & -0.10 & -0.03  &-0.11 & -0.02 & -0.15 & -0.03  & -0.15 & -0.03   &-0.11&-0.04\\
  3 & 0.17 & 0.06  & 0.18 & 0.08 &  0.14 & 0.06  & 0.13 & 0.05           &0.13&0.02\\
  4 & 0.07 & 0.01  & 0.08 & 0.02 &  0.04 & 0.01  & 0.04 & 0.00             &0.03&0.00\\
  5 & -0.05 & 0.02 & -0.06  & 0.04 & -0.10  & 0.03  & -0.10 & 0.02      & &\\
  6 & 0.01 & 0.03  & 0.04 & 0.04 &  0.00  & 0.02  & 0.00 & 0.02         & & \\
  7 & 0.01 & -0.01  & 0.04 & 0.01 &  0.01 & -0.01  & 0.01 & -0.01     & &\\
\end{tabular}\\
\leftline{\textsuperscript{\emph{a}}ref~\onlinecite{Muscat1999}.}
\end{table*}

\begin{table}
\caption{The dependence of surface energy of TiO$_{2}$(100) on slab thickness, in unit of J/m$^2$.}
\label{tbl:4}
\begin{tabular}{l c c c c c c c }
    Layer  & 5   & 9 & 13  & 17 & Ref.\textsuperscript{\emph{a}} &  Ref.\textsuperscript{\emph{b}} & Ref.\textsuperscript{\emph{c}} \\\hline
  $\gamma$ & 0.70  & 0.68   & 0.68  & 0.68 & 0.67 & 0.69 & 0.83 \\
\end{tabular}\\
\leftline{\textsuperscript{\emph{a}} ref~\onlinecite{Perron2007};
\textsuperscript{\emph{b}} ref~\onlinecite{Labat2008};
\textsuperscript{\emph{c}} ref~\onlinecite{Muscat1999}.}
\end{table}

\section{RESULTS: Interface properties}

The orientation relationship is set as Ti(10$\overline{1}$0)[0001] //
TiO$_{2}$(100)[010] across the interface. The size of the
Ti(10$\overline{1}$0) surface cell is 2.934 \AA~ $\times$ 4.638 \AA,
while it is 2.971 \AA~ $\times$ 4.645 \AA~ for TiO$_{2}$(100). A
coherent interface is obtained by a small strain of the softer Ti to
perfectly match the TiO$_{2}$(100) surface cell: this has little
effect on the energy since the mismatch of the two surface cells is
so small. To check dependence on boundary conditions we used two
models to simulate the interface, a film-like model with an interface
and two free surfaces, and a bulk-like model with two interfaces. The
interface structures were relaxed with k-point mesh $3\times3\times1$
(four irreducible k points).

\subsection{Film-like interface model}

The film-like model is generated with 16-layers of Ti, 13-layers of TiO$_{2}$ and a 15 \AA~ vacuum region. This gives one interface and two free surfaces.

Various terminations of the surfaces and metal/oxide interface are
possible, their stability depending on the environmental condition
like the partial pressure of O$_{2}$ gas.\cite{Zhang2000} As we
mentioned above, TiO$_{2}$(100) has three possible terminations, with
uppermost layers O-O-Ti, O-Ti-O, and Ti-O-O.  For interfaces, we
considered different stacking sequences of Ti onto these terminations.
Following work on other interfaces\cite{Siegel2002, Liu2003},
the first layer of the Ti-metal is placed...
\begin{itemize}
\item
 \texttt'OT\texttt': directly above uppermost Ti cations;
\item \texttt'HCP\texttt':  above the second
layer of  Ti cations;
\item \texttt'FCC\texttt': above the third layer of Ti cations.
\item \texttt'TT\texttt': for O-O-Ti, as an extension of the oxide.
\item \texttt'TT\texttt': for Ti-O-O, as an extension of the metal.
\end{itemize}

The \texttt'TT\texttt' configurations are those which would allow growth of the
oxide/metal by simple extension of a stable interface, with the Ti
\texttt'atoms\texttt' and Ti \texttt'cations\texttt' directly adjacent. In fact, the
two \texttt'TT\texttt' configurations are the same at the interface, the calculation
differing in the accompanying surface.

Despite this apparently exhaustive survey of stacking
sequences, we also found that where the TiO$_{2}$(100) surface is
quite corrugated, significant reconstruction can occur within the Ti-metal
region, spontaneously generating a stacking fault.
When this happens, we
report the most stable relaxed structure, and label it
\texttt'HCP-2\texttt' and \texttt'FCC-2\texttt' in \ref{tbl:wad}.

To avoid spurious contraction due to surface tension, we fix the
lattice parameters in the interface plane. But we allow relaxation perpendicular by first calculating the total energy of
the unrelaxed interface structure as a function of interface separations $d$. Once the optimal value is found, all the atoms are relaxed
to the ground state at fixed volume.

In order to evaluate the strength of the interface, we calculated the work of adhesion, $W_{ad}$, which is defined as:

\[W_{ad} = (E^{\rm{tot}}_{\rm{M}} + E^{\rm{tot}}_{\rm{O}} - E^{\rm{tot}}_{\rm{M/O}})/S \]
where $E^{\rm{tot}}_{\rm{M}}$ is the surface energy of Ti-metal, $E^{\rm{tot}}_{\rm{O}}$ is the surface energy of oxide, $E^{\rm{tot}}_{\rm{M/O}}$ is the energy of the interface structure, and S is the interface area.
A positive value of $W_{ad}$ means that the interface is energetically favorable over the free surfaces. \ref{tbl:wad}  shows that  $W_{ad}$
increases with increasing number of interfacial O atoms (Ti-O-O,
O-Ti-O, and O-O-Ti).  However, much of this is due to
 differences in surface stability;
e.g.  the O-O-Ti and Ti-O-O with \texttt'TT\texttt' stacking
have the same
chemical composition near the interface but
very different $W_{ad}$.

To understand the growth of an additional TiO$_2$ layer, each of these interface types needs to be considered. In the following subsections, the interface with the most stable stacking sequence for each termination is analyzed in detail.

\begin{table*}
\caption{Relaxed values of $W_{ad}$ (J/m$^2$) for interfaces with different termination and stacking sequence, with interfacial distance $d$ (\AA).}
\label{tbl:wad}
\begin{tabular}{l c c c c c | c c c c | c c c c}
           &  \multicolumn{5}{c}{O-O-Ti} & \multicolumn{4}{c}{O-Ti-O}&  \multicolumn{4}{c}{Ti-O-O}  \\
           & OT & HCP & FCC & TT & HCP-2        & OT  & HCP & FCC & FCC-2      & OT  & HCP & FCC & TT \\ \hline
  d         & 1.65 & 0.57 & 1.09 & 0.97 & 1.17   & 1.87 & 1.37 & 0.33 & 0.31 & 1.90 & 1.66 & 1.58 & 0.55 \\
  $W_{ad}$  & 7.54 & 9.43 & 8.99 & 9.32 & 10.34  & 2.28 & 2.57 & 3.49 & 3.55 & 1.41 & 2.59 & 2.04 & 2.83 \\
\end{tabular}
\end{table*}

For all configurations listed in \ref{tbl:bader}, for the
metal slab, the Ti atom in the center has a charge 0.00, as in bulk
Ti, which again confirms that the slab we used is thick enough.  For
the TiO$_{2}$ slab, the charge associated with central Ti and O ions
are 2.18\emph{e} and -1.10\emph{e}, respectively. This agrees well
with DFT$+U$ results from Jess and the co-workers\cite{Jess2010} who
found the Bader charge is 2.22\emph{e} and -1.12\emph{e} for Ti and
O. Although Bader charge values are far from the formal charge
4\emph{e} (Ti$^{4+}$) and -2\emph{e} (O$^{2-}$), we believe that
useful bonding information can be obtained from the charge transfer.

\subsubsection{Atomic, electronic structure of O-O-Ti interface}

\begin{figure}
\includegraphics[width=7cm]{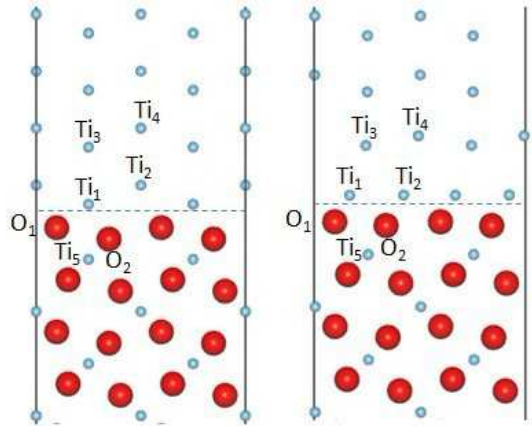}
\caption{\texttt'HCP-2\texttt' structure of O-O-Ti interface (only some layers near the interface is shown), viewed along [001] direction of TiO$_{2}$, the dashed line indicates the position of the interface. Left: without atomic relaxation (the equivalent atoms on the boundary are also shown). Right: with atomic relaxation. Large red spheres are O atoms, and small silver spheres are Ti atoms.\label{fig:3}}
\end{figure}

The stacking sequence \texttt'HCP-2\texttt' gives rise to the most stable O-O-Ti interface (\ref{tbl:wad}). The structure is displayed in \ref{fig:3}. To observe the geometry change due to the atomic relaxation clearly, the unrelaxed (input) structure is also shown and the ions with large displacement are labeled. On the metallic side, Ti$_{2}$ and Ti$_{4}$ move towards the oxide. Ti$_{1}$ and Ti$_{2}$ are found shifting along [000$\overline{1}$] of Ti significantly. As a result, the surface morphology of Ti(10$\overline{1}$0) is totally changed: the Ti$_{1}$ and Ti$_{2}$ layers merge to give a flat surface; the spacing between Ti$_{2}$ and Ti$_{3}$ layer changes to 2.19 \AA~ from 1.69 \AA~ in the bulk; the interlayer spacing between Ti$_{3}$ and Ti$_{4}$ decreases to 0.43 \AA~ from the bulk spacing 0.85 \AA. For the oxide slab, the relaxation is less significant, the interfacial atoms move towards metal slab, in particularly O$_{2}$, thus the surface of the oxide becomes flatter. Across the interface, the separations of Ti$_{1}$-O$_{1}$ and Ti$_{2}$-O$_{2}$ is 2.00 \AA~ and 2.15 \AA, which is close to the Ti-O bond lengths in bulk TiO$_{2}$ (1.96 \AA~ and 2.00 \AA). O$_{2}$ is located in the center of the \texttt'half\texttt' octahedron (formed by Ti$_{1}$, Ti$_{2}$ and Ti$_{5}$), and O$_{1}$ sits in a similar octahedral site.

Bader charge analysis\cite{Tang2009} is applied to study the charge transfer. In \ref{tbl:bader}, we give the Bader charge associated with the labeled atoms (\ref{fig:3}) as well as the atoms in the center of the metal/oxide slabs, and the value of the charge is the variation from the neutral Ti/O atom (10\emph{e} for Ti, and 6\emph{e} for O). Thus a negative value means accepting negatively charged electrons, while a positive value means donating electrons.

\begin{table*}
\caption{Bader charge (\emph{e}) of the selected atoms in the various interface structures (denoted by the termination, the model used, and the staking sequence in the bracket). Definitions of the atoms are given in the appropriate figures. Bulk values for Bader charge are 0.00 in Ti, and 2.18$e$ and -1.10$e$ for Ti and O in bulk rutile. For all the structures listed in the table, the Ti atom/ion in the center of the slab has the same Bader charge as in the bulk.}
\label{tbl:bader}
\begin{tabular}{l c c c c | c c  c c}
                &   \multicolumn{4}{c}{Ti slab} &  \multicolumn{4}{c}{TiO$_{2}$ slab} \\
 Atom           & Ti$_{1}$ & Ti$_{2}$ & Ti$_{3}$ & Ti$_{4}$   & Ti$_{5}$ & O$_{1}$ & O$_{2}$  & O(center)\\\cline{2-5} \cline{6-9}
O-O-Ti film(HCP-2) & 0.81    & 0.82  & -0.21  &  0.32        &  2.08      &  -1.20   &  -1.28    &  -1.11   \\
O-Ti-O film(FCC-2)& 0.82 &  0.06 & 0.24  & 0.01             &  1.78       &  -1.37   &  -1.24    &  -1.10 \\
Ti-O-O film(TT) & 0.56  &  -0.06  & 0.18 & -0.04            &  1.12    &  -1.27   &  -1.20    &  -1.09 \\
S-O\textsuperscript{\emph{a}} bulk(FCC) & 0.77    & 0.15  & 0.23  &  -0.09    &  1.78     &  -1.39   &  -1.24    &  -1.11   \\
D-O\textsuperscript{\emph{b}} symmetry bulk(TT) & 1.05  &  0.62   & -0.01    &  0.12   &  2.10    &  -1.21   &  -1.27    &  -1.11 \\
D-O\textsuperscript{\emph{b}} asymmetry bulk(TT)  & 1.06  &  0.60   & -0.05    &  0.15   &  2.11     &  -1.27   &  -1.21    &  -1.11 \\
D-O\textsuperscript{\emph{b}} asymmetry bulk(HCP-2) & 0.83    & 0.81  & -0.19  &  0.31    &  2.08    &  -1.19   &  -1.28    &  -1.11 \\
\end{tabular}
\leftline{\textsuperscript{\emph{a}} Single-Oxygen;
\textsuperscript{\emph{b}} Double-oxygen.}
\end{table*}

For \texttt'HCP-2\texttt' each of the interfacial atoms Ti$_{1}$ and Ti$_{2}$ donates
0.8 electrons, which account for the net charge transfer of 1.61
electrons from the metal slab to the oxide (\ref{tbl:bader}). Bader analysis gives a negative charge for
Ti$_{3}$, but this appears to be an artifact due to the large
relaxation around it, leading to a large Bader volume, which in turn
encloses more electrons.  It is better to consider the layer
containing Ti$_3$ and Ti$_4$, which has overall a small positive
charge.  The interfacial oxygen ions O$_{1}$ and
O$_{2}$ attract more electronic charge than the oxygen in the center,
due to the more electron donators nearby. Ti$_{5}$ transfers fewer
electrons than the Ti ions in the center of the oxide, which is also
understandable as its coordination is smaller. The large electron
transfer found here implies strong ionic bonding across the interface.

\begin{figure}
\includegraphics[width=8cm]{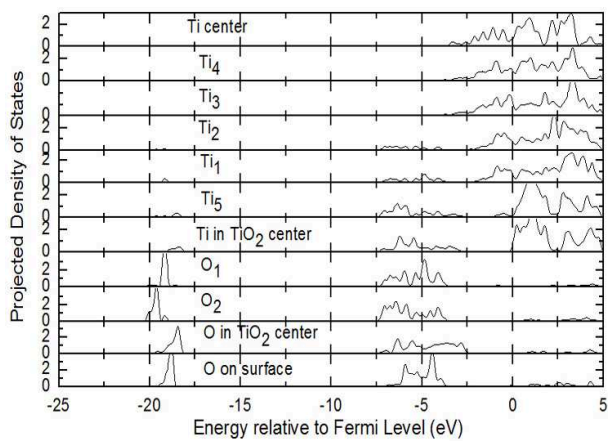}
\caption{DOS projected onto selected atoms for the O-O-Ti interface. The labels of the atoms are identified in \ref{fig:3}. Note that
within TiO$_{2}$ all the states corresponding to the 4s3d band are unoccupied, consistent with formal ionic charges of Ti$^{4+}$ and O$^{2-}$.\label{fig:4}}
\end{figure}

\begin{figure}
\includegraphics[width=8cm]{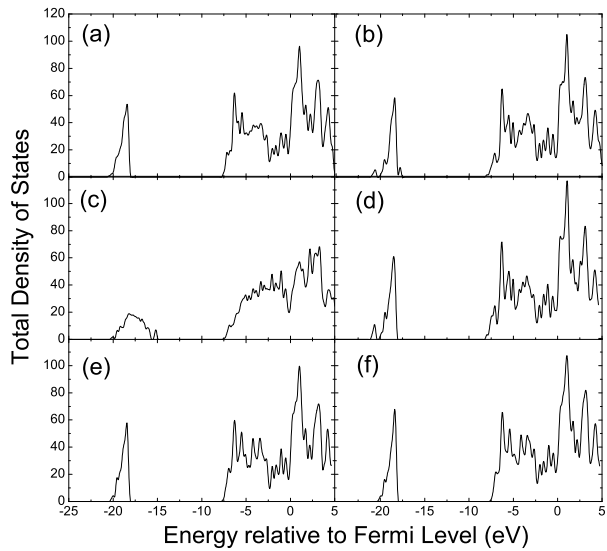}
\caption{Total DOS of various interfaces using two models. (a) Total DOS of the O-O-Ti terminated interface using film-like model. (b) Total DOS of the O-Ti-O terminated interface using film-like model. (c) Total DOS of the Ti-O-O terminated interface using film-like model. In this interface slab, a O-O-Ti terminated TiO$_{2}$(100) free surface is included. This polarized surface might account for the abnormal shape of the 2s states of oxygen. (d) Total DOS of the single-oxygen interface using bulk-like model. (e) Total DOS of the double-oxygen interface using symmetric structure of bulk-like model (including two \texttt'TT\texttt' stacking interfaces). (f) Total DOS of the double-oxygen interface using asymmetric structure of bulk-like model (including one \texttt'TT\texttt' and one \texttt'HCP-2\texttt' stacking interface). \label{fig:5}}
\end{figure}

The electronic density of states is projected onto selected atoms to determine the bonding character (\ref{fig:4}).
Notably, Ti$_{3}$ and Ti$_{4}$ are dissimilar, as is reflected in the peak/valley just below the Fermi level.
The interfacial atoms Ti$_{1}$ and Ti$_{2}$ show a small hybridization peak in the region from -7.5 eV to -2.5 eV, representing weak covalent bonding to the oxygen nearby. Strong hybridization between the Ti $3d$ and O $2p$ is observed from the DOS of all ions on the TiO$_{2}$ side of the interface, indicating some covalent bonding in rutile.\cite{Paxton1998} Ti$_{5}$ has almost no Ti  conduction band density: which implies that bonding in the O-O-Ti structure is not metallic. The DOS of O$_{1}$ and O$_{2}$ are similar to the oxygen in the center of oxide slab, with a small shift of both s and p peaks to a lower energy level indicating a stronger Madelung field at the interface, which can stabilize the system.\cite{Landa2009} For the oxygen on the surface, the s orbital shows no shift, and the width of its p orbital is reduced. By comparing the total DOS of the interface slab (\ref{fig:5}(a)) and DOS of pure TiO$_{2}$ (\ref{fig:1}), we see a small peak around -7.5 eV induced by the interface atoms. This will be discussed further later.

\subsubsection{Atomic, electronic structure of O-Ti-O interface}

\begin{figure}
\includegraphics[width=7cm]{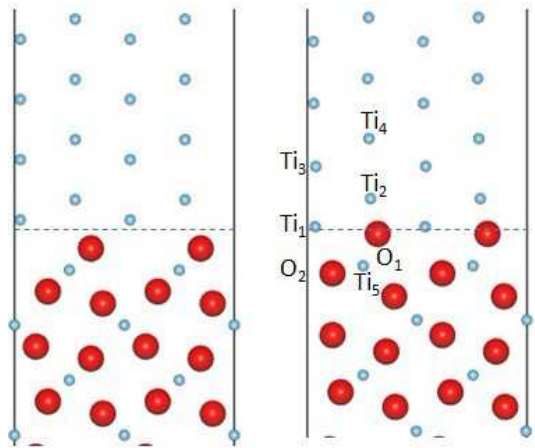}
\caption{\texttt'FCC-2\texttt' structure of O-Ti-O interface (only several layers near the interface is shown), viewed along [001] direction of TiO$_{2}$, the dashed line indicate the position of the interface. Left: without atomic relaxation (the equivalent atoms on the boundary are also shown). Right: with atomic relaxation. Red spheres are O atoms, and silver spheres are Ti atoms.\label{fig:6}}
\end{figure}

As shown in \ref{tbl:wad}, for O-Ti-O interface systems, the interface with strongest adhesion is \texttt'FCC-2\texttt', which has a similar structure to \texttt'FCC\texttt'. \ref{fig:6} shows the optimized \texttt'FCC-2\texttt' interface: Ti$_{1}$ and Ti$_{3}$ are shifted significantly toward the oxide, forming planes containing both Ti and O atoms. With this displacement, the interlayer spacing of the metal slab near the interface changed to 1.19 \AA~ and 1.36 \AA, which are 0.85 \AA~ and 1.69 \AA~ in the bulk Ti, respectively. Relaxation of O$_{1}$ and O$_{2}$ leads to a displacement towards the metal even larger than that in the free surface structure. Finally, the separation O$_{1}$-Ti$_{2}$ and O$_{2}$-Ti$_{1}$ is 2.13 \AA~ and 2.10 \AA,  close to the spacing in bulk TiO$_{2}$ (1.96/2.00 \AA).

The Bader charge analysis results are listed in \ref{tbl:bader}.
The most significant charge transfer comes from Ti$_{1}$
and Ti$_{3}$, which donate a total
 of 1.16 electrons from the metal to the oxide. The excess charge is located on the interfacial atoms (Ti$_5$, O$_{1}$ and O$_{2}$).

\begin{figure}
\includegraphics[width=8cm]{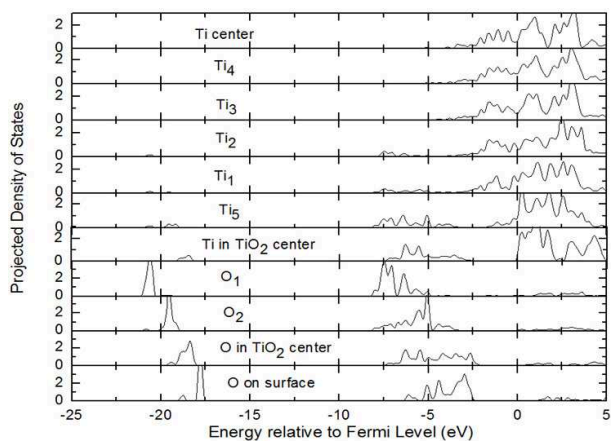}
\caption{DOS projected onto selected atoms for the O-Ti-O interface. The label of the atom is identified in \ref{fig:6}.\label{fig:7}}
\end{figure}

As seen from \ref{fig:7}, the atom-projected DOS converges rapidly to bulklike values
away from the interface.  For the interfacial ion Ti$_{5}$,
states in the band gap of the oxide implies a kind of
metallic behavior, and it is a little stronger than that in O-O-Ti
interface, because of its direct exposure to the metal slab. The shift
of DOS is very apparent in the interfacial atoms O$_{1}$ and O$_{2}$,
especially for O$_{1}$, the 2s states is below -20 eV. For the surface
oxygen, the 2s and 2p orbitals move to a higher energy level, which
contribute to the small peak near -17.5 eV in the total DOS
(\ref{fig:5}(b)). Also, comparing \ref{fig:1} (pure TiO$_{2}$),
\ref{fig:5}(b) (interface with surface), and \ref{fig:5}(d)
(interface with no surface), it can be concluded that,
in \ref{fig:5}(b), the small peak below -20 eV and peak near -7.5 eV are
associated with the interface atoms, while the peak near -17.5 eV comes from
surface atoms. These agree with the our calculation results that the
formation of the interface is exothermic, while that is endothermic
for the surface.

\subsubsection{Atomic, electronic structure of Ti-O-O interface}

\begin{figure}
\includegraphics[width=7cm]{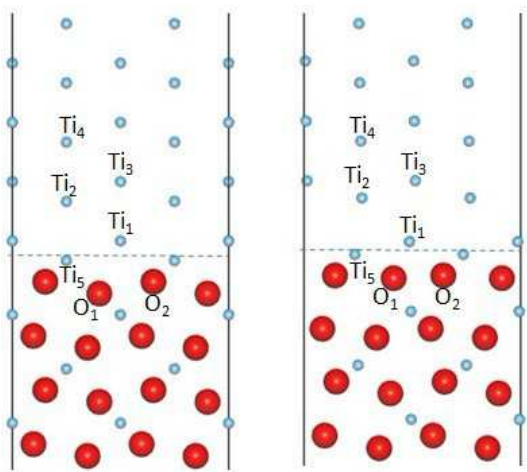}
\caption{\texttt'TT\texttt' structure of Ti-O-O interface (only several layers near the interface is shown), viewed along [001] direction of TiO$_{2}$, the dashed line indicate the position of the interface. Left: without atomic relaxation (the equivalent atoms on the boundary are also shown). Right: with atomic relaxation. Red spheres are O atoms, and silver spheres are Ti atoms.\label{fig:8}}
\end{figure}

For the Ti-O-O structures, \texttt'TT\texttt' stacking
exhibits the lowest energy. The input and relaxed structure of
\texttt'TT\texttt' stacking are shown in
\ref{fig:8}. Again, the relaxation tends to flatten the surfaces. The
interfacial spacing (distance between Ti$_{1}$ layer and Ti$_{5}$
layer) is 0.55 \AA, smaller than the 0.85 \AA~ of the interlayer
spacing in bulk Ti. O$_{1}$ moves towards the interface, which
increases the effective coordination of Ti$_{5}$. The downward shift
of Ti$_{1}$ and Ti$_{3}$ are also observed. The interlayer spacing of
Ti$_{1}$-Ti$_{2}$ and Ti$_{2}$-Ti$_{3}$ is 1.88 \AA~ and 0.74 \AA,
respectively, which slightly deviate from the spacing in the bulk Ti.

\begin{figure}
\includegraphics[width=8cm]{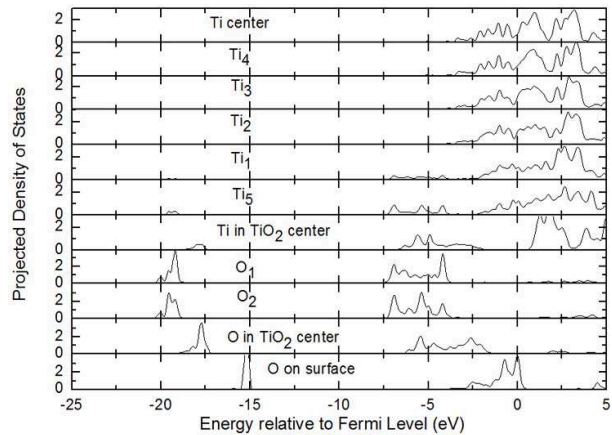}
\caption{DOS projected onto selected atoms for the Ti-O-O interface. The label of the atom is identified in \ref{fig:8}.\label{fig:9}}
\end{figure}

The Bader charge analysis (\ref{tbl:bader}) and DOS projection (\ref{fig:9}) are qualitatively similar to the two systems. However, only 0.41\emph{e} is moved to the oxide slab from the metal, much less than in the other two surfaces. From the projected DOS of the interfacial ions (Ti$_{1}$,Ti$_{2}$, Ti$_{3}$ and Ti$_{4}$), no covalent features are evident between the two slabs, but metallic bonding can be inferred from the occupied 4s3d states in Ti$_{5}$ which is adjacent to the metal slab. The relatively small $W_{ad}$ of the Ti-O-O systems can be understood by the pDOS (\ref{fig:9}), where the 2s states of surface oxygen move to a very high level just below -15 eV. Thus the states between -15 eV and -17.5 eV in the \ref{fig:5}(c) come from the oxygen near the free surface (O-O-Ti terminated) included in the slab model.

\subsection{Bulk-like interface model}

In this model, we consider a periodic arrangement $\cdots$metal-oxide-metal$\cdots$, with no free surfaces.
One advantage of this model over the previous one is that the effects of surface dipole
interactions are eliminated.

Considering the termination of TiO$_{2}$(100) at the interface, there are two structures with different
layer ordering
\begin{itemize}
\item $\cdots$Ti-Ti-Ti-O-Ti-O-O-Ti$\cdots$O-O-Ti-O-Ti-Ti-Ti$\cdots$,
\item $\cdots$Ti-Ti-Ti-O-O-Ti-O-O-Ti$\cdots$O-O-Ti-O-O-Ti-Ti-Ti$\cdots$.
\end{itemize}
The former, denoted as {\it single-oxygen interface}, may have either two identical interfaces or different stacking sequences.
For the latter,  {\it double-oxygen interface}, we first calculate a \texttt'TT\texttt' configuration where the first cation (metal) layer is like the extension of the metal (oxide). Subsequently we made an \texttt'asymmetric\texttt' structure: one interface with \texttt'TT\texttt' stacking, the other with \texttt'HCP-2\texttt' stacking, so that the energy of the double-oxygen interface with \texttt'HCP-2\texttt' stacking can be obtained. Again, we fix the dimensions of the cell in the interface plane.

In order to evaluate the strength of the interface, we defined the interface energy $W_{bulk}$ as:

\[ W_{bulk} = (E^{\rm{int}}_{\rm{TimOn}} - \frac{1}{2}E^{\rm{bulk}}_{\rm{TiO2}} - (m-\frac{n}{2})E^{\rm{bulk}}_{\rm{Ti}})/2S \]
where $E^{\rm{int}}_{\rm{TimOn}}$ is the energy of interface structure consisting of m Ti atoms and n O atoms. $E^{\rm{bulk}}_{\rm{TiO2}}$ is the energy per TiO$_{2}$ unit in the bulk oxide, $E^{\rm{bulk}}_{\rm{Ti}}$ is the energy per Ti atom in the bulk metal, and 2S is the area of the two interfaces in the interface structure.

The results are summarized in \ref{tbl:X}. The negative values indicate that the interface is favored relative to the bulk.
In contrast to the results for the O-O-Ti terminated interface shown in \ref{tbl:wad}, \texttt'HCP-2\texttt' stacking here is less stable than the \texttt'TT\texttt' stacking by about 0.1 J/m$^2$.

\begin{figure}
\includegraphics[width=7cm]{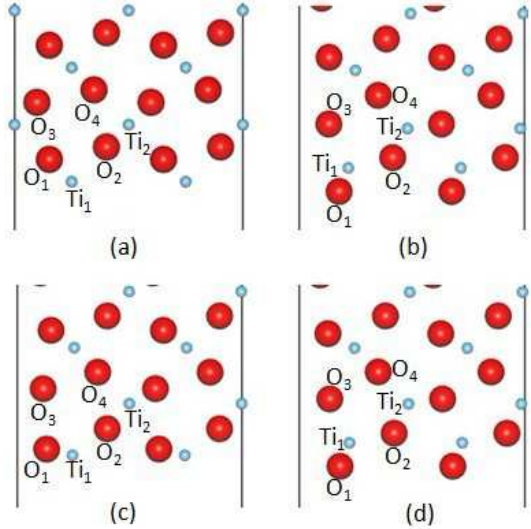}
\caption{Structure of the Ti-O-O terminated TiO$_{2}$(100) surface. (a) unrelaxed structure (bulk termination), (b) relaxed structure in the TiO$_{2}$(100) surface slab, (c) relaxed structure in the O-O-Ti terminated interface slab with \texttt'TT\texttt' stacking, (d) relaxed structure in the O-O-Ti terminated interface slab with \texttt'HCP-2\texttt' stacking. Red spheres are O atoms, and silver spheres are Ti atoms.\label{fig:10}}
\end{figure}

 To illustrate the possible effect of the free surface on the $W_{ad}$
 obtained by the film-like model, in \ref{fig:10}, we display the
 structure of the Ti-O-O free surface from the film-like
 interface slab with \texttt'TT\texttt' and
 \texttt'HCP-2\texttt' stacking, compared with its
 structure in the pure TiO$_{2}$(100) calculation. After
 relaxation in the surface slab (\ref{fig:10}(b)),
 O$_{1}$ and O$_{2}$ move significantly outwards, and
 finally, the Ti-O-O termination transforms to an arrangement like
 O-Ti-O. In TiO$_{2}$, the surface terminated with Ti is
 highly polar and this outward movement of oxygen can decrease the
 dipolar moment. Very similar relaxation is observed in the interface
 with \texttt'HCP-2\texttt' stacking
 (\ref{fig:10}(d)). However, for the \texttt'TT\texttt'
 stacking (\ref{fig:10}(c)), the relaxation of O$_{1}$ and O$_{2}$, does not occur,
 and the termination remains Ti-O-O. As we mentioned
 above, in \texttt'TT\texttt' stacking, the first Ti
 layer of the metal sits in a position like the extension of the
 oxide, so the O-O-Ti terminated interface with a stoichiometric
 oxide can also be seen as a Ti-O-O terminated interface with a
 non-stoichiometric oxide. This different surface relaxation
 directly affects the work of adhesion of the
 interface. We do not think this kind of interaction between the
 interface and polarized surface can be removed by increasing the
 thickness of the oxide slab.

\begin{table*}
\caption{Bader charge (\emph{e}) of the selected atoms on Ti-O-O surface included in different structures, the labels of the atoms are identified in \ref{fig:10}(b), (c), and (d). Bulk values for Bader charge 2.18$e$ and -1.10$e$ for Ti and O in bulk rutile.}
\label{tbl:polar}
\begin{tabular}{l c c c | c c c}
                &   \multicolumn{3}{c}{surface TiO$_{2}$ unit} &  \multicolumn{3}{c}{inside TiO$_{2}$ unit} \\ \cline{2-4} \cline{5-7}
 Atom           & Ti$_{1}$ & O$_{1}$ & O$_{2}$ & Ti$_{2}$ & O$_{3}$ & O$_{4}$ \\
 free surface   & 1.94    &  -1.03   & -1.19    & 1.88  & -1.23      &   -1.13  \\
 TT stacking    & 1.56    &  -1.20  & -1.16     & 2.11   & -1.14      &  -1.10   \\
 HCP-2 stacking & 1.93    &  -1.04  & -1.19     & 1.88   & -1.25      &  -1.14   \\
\end{tabular}
\end{table*}

In order to analyze the effects of interfaces on the Ti-O-O surface in detail, we examine the Bader charge on
the ions near the surface in \ref{tbl:polar}.
In all cases the electron density  near the surface is enhanced by almost one electron per surface cell.
The charge distribution near the free surface is unaffected by an
\texttt'HCP-2\texttt' interface, but the presence of a
\texttt'TT\texttt' interface, changes the surface charge distribution significantly.
This interaction between surface and interface cast doubt on the validity of the film-like model.
Similar calculations for the single-oxygen structures (terminated with O-Ti-O), show
no such  structure or Bader charge variation. Thus for the double-oxygen interface, the film-like model is still reliable.

So get an estimate of finite size error, we can directly compare the
calculated work of adhesion for the same interface obtained by
bulk-like and film-like calculation (\ref{tbl:wad}),and
this requires subtraction of the
energy of the free surfaces. These
are: 1.99 J/m$^2$ and  2.20 J/m$^2$
for Ti(10$\overline{1}$0) terminated with small
and large interlayer spacing respectively;
0.68 J/m$^2$ for TiO$_{2}$(100)
terminated with O-Ti-O, and 7.91 J/m$^2$ for the sum of Ti-O-O and
O-O-Ti terminations.
The resultant interface
energy $W_{film}$ is listed in \ref{tbl:X}. For the
single-oxygen interfaces, $W_{bulk}$ is in reasonable agreement with
$W_{film}$. For the double-oxygen interface with
\texttt'TT\texttt' stacking the apparent discrepancy is obviously
due to the different surface reconstruction, as just described.
And the bulk-like model is more reasonable.
In the next  subsections, we will discuss the geometry as
well as the electronic structure difference for these two models.

\begin{table}
\caption{Relaxed values for $W_{bulk}$ (J/m$^2$), compared with $W_{film}$ that obtained by excluding the contribution of the surface energy from the work of adhesion by the film-like model.}
\label{tbl:X}
\begin{tabular}{l c c c c | c c}
  Termination & \multicolumn{4}{c}{single-oxygen}&  \multicolumn{2}{c}{double-oxygen}\\
  Stacking    & OT & HCP & FCC & FCC-2       & TT & HCP-2\\ \hline
  $W_{bulk}$  & 0.30 & -0.03 & -0.95 & -0.93 & -0.60 & -0.51\\
  $W_{film}$  & 0.39 & 0.10  & -0.82 & -0.88 & -0.05 & \\
\end{tabular}
\end{table}

\subsubsection{Atomic, electronic structure of single-oxygen interface}

\begin{figure}
\includegraphics[width=7cm]{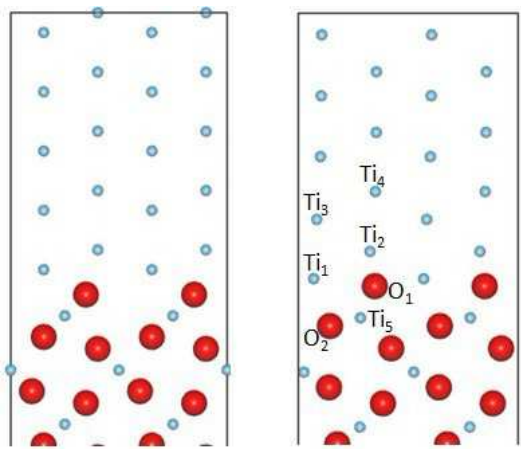}
\caption{Structure of single-oxygen interface with \texttt'FCC\texttt' stacking, viewed along [001] direction of TiO$_{2}$. Only layers near one interface are shown: the other is symmetrically equivalent. Left: without atomic relaxation (the equivalent atoms on the boundary are also shown). Right: with atomic relaxation. Red spheres are O atoms, and silver spheres are Ti atoms.\label{fig:11}}
\end{figure}

\begin{figure}
\includegraphics[width=8cm]{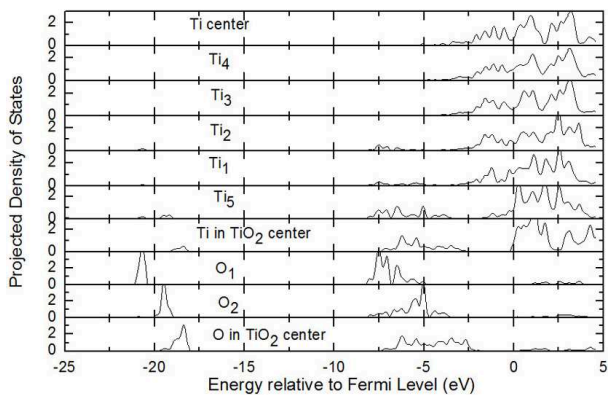}
\caption{DOS projected onto selected atoms for the single-oxygen interface. The label of the atom is identified in \ref{fig:11}.\label{fig:12}}
\end{figure}

The structure of \texttt'FCC\texttt' (\ref{fig:11}) is quite similar as the \texttt'FCC-2\texttt' (\ref{fig:6}), in particularly the configuration near the interface and relaxations of the interfacial atoms (Ti$_{1}$, Ti$_{3}$, O$_{1}$, O$_{2}$) in the relaxed structure. Atom spacing O$_{1}$-Ti$_{2}$ and O$_{2}$-Ti$_{1}$ is 2.10 \AA~ and 2.12 \AA, very close to the results in \texttt'FCC-2\texttt' stacking obtained by the film-like model, the Bader charge (\ref{tbl:bader}) and projected DOS on the interfacial atoms (\ref{fig:12}) are also similar to the film-like model. For the total DOS (\ref{fig:5}(d)), the surface-atom-induced small peak around -17.5 eV is absent, which is expected.

\subsubsection{Atomic, electronic structure of double-oxygen interface using symmetric structure}

\begin{figure}
\includegraphics[width=7cm]{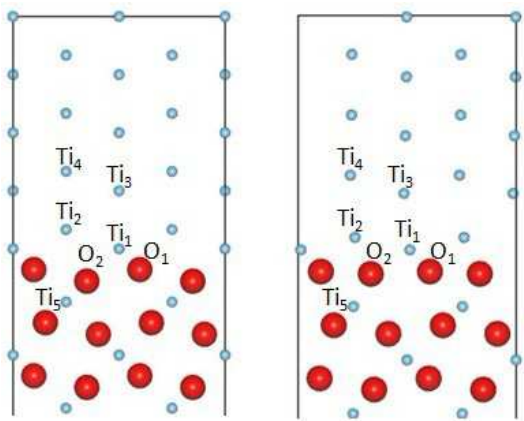}
\caption{Structure of double-oxygen interface with \texttt'TT\texttt' stacking (due to the symmetry, only layers near one interface is shown), viewed along [001] direction of TiO$_{2}$. Left: without atomic relaxation (the equivalent atoms on the boundary are also shown). Right: with atomic relaxation. Red spheres are O atoms, and silver spheres are Ti atoms.\label{fig:13}}
\end{figure}

Not surprisingly, \ref{fig:13} is very similar to the \texttt'TT\texttt' stacking of Ti-O-O terminated interface obtained using the film-like model. For example, the layer spacing of Ti$_{1}$-Ti$_{2}$ is 0.54 \AA, corresponding to 0.55 \AA~ for Ti$_{1}$- Ti$_{5}$ layer in \ref{fig:8}; while the spacing of Ti$_{2}$-Ti$_{3}$ is 1.90 \AA, very close to the Ti$_{1}$-Ti$_{2}$ spacing 1.88 \AA~ (\ref{fig:8}).  Note that
Ti$_{1}$ in \ref{fig:13} is in the equivalent position to Ti$_{5}$ in
\ref{fig:8}.

\begin{figure}
\includegraphics[width=8cm]{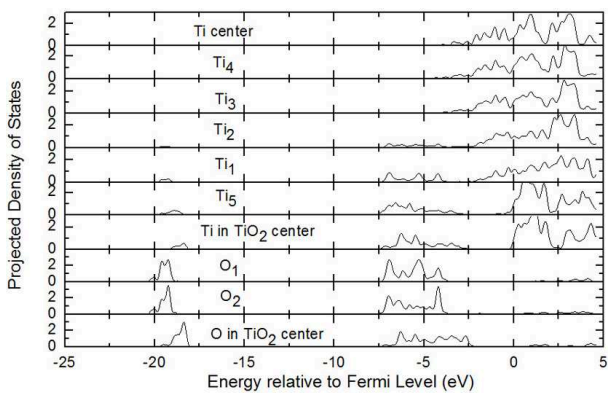}
\caption{DOS projected onto selected atoms for the double-oxygen interface using symmetric structure. The label of the atom is identified in \ref{fig:13}.\label{fig:14}}
\end{figure}

Again the Bader charge (\ref{tbl:bader}) and projected DOS (\ref{fig:14}) agree well with the film-like model (\ref{fig:9}).
Comparing the total density of states (\ref{fig:5}(c) and \ref{fig:5}(e)),
only the surface oxygen states in the range from -17.5 eV to -15 eV
are absent.

\subsubsection{Atomic, electronic structure of double-oxygen interface using asymmetric structure}

\begin{figure}
\includegraphics[width=8cm]{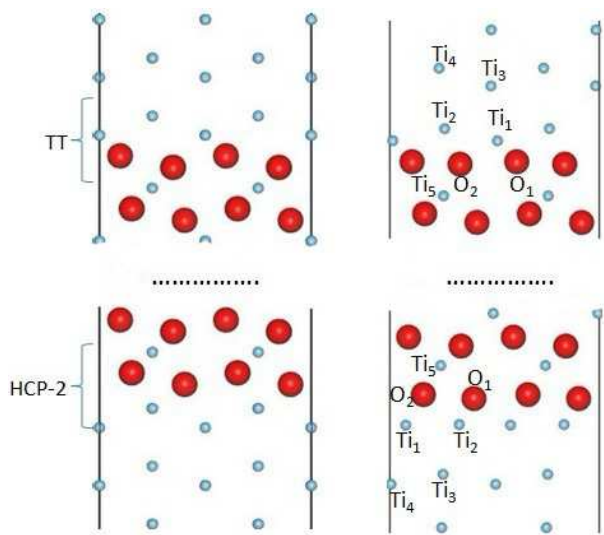}
\caption{Structure of two double-oxygen interfaces with different stacking sequence, viewed along [001] direction of TiO$_{2}$. The ions in the center of the oxide are omitted. Left: without atomic relaxation (the equivalent atoms on the boundary are also shown). Right: with atomic relaxation. Red spheres are O atoms, and silver spheres are Ti atoms.\label{fig:15}}
\end{figure}

\begin{figure}
\includegraphics[width=8cm]{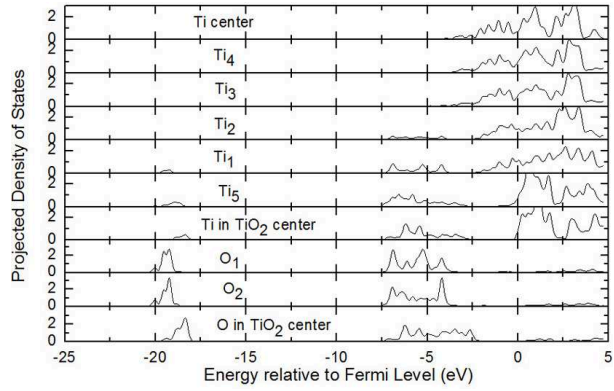}
\caption{DOS projected onto selected atoms for the \texttt'TT\texttt' stacking interface. The label of the atom is identified in \ref{fig:15}.\label{fig:16}}
\end{figure}

\begin{figure}
\includegraphics[width=8cm]{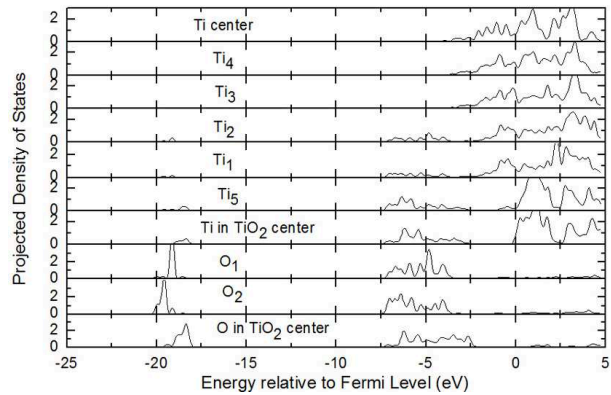}
\caption{DOS projected onto selected atoms for the \texttt'HCP-2\texttt' stacking interface. The label of the atom is identified in \ref{fig:15}.\label{fig:17}}
\end{figure}

As shown in \ref{fig:15}, we can consistently reproduce the interface structure with \texttt'TT\texttt' stacking (\ref{fig:8} and \ref{fig:13}) and \texttt'HCP-2\texttt' stacking (\ref{fig:3}) very well in film, symmetric-bulk and asymmetric-bulk calculation. The Bader charges ( \ref{tbl:bader}) are equivalent to film-like model. The relevant projected DOS compare very well with each other (\ref{fig:16} vs \ref{fig:14} , and \ref{fig:17} vs \ref{fig:4}). Again no surface-atom-induced peak is found in the total DOS figure (\ref{fig:5}(f)).  We can therefore be highly confident that these structures  are converged with cell size and not artifacts of the boundary conditions.

\section{CONCLUSION}

First-principles calculation is performed to investigate the atomic
structure and bonding nature of the Ti(10$\overline{1}$0) //
TiO$_{2}$(100)interface. A series of tests have been done to verify the
settings used in the calculation, by studying the bulk and free
surface properties and successful comparison with previous calculation
and experimental work.

Two different models are used to simulate the interface. With the
film-like model, we considered three terminations and different
stacking sequences.  Interfaces compatible with the extension
mechanism (\texttt'TT\texttt' stacking) were
constructed for the O-O-Ti and Ti-O-O systems, and we found these
geometries quite favored, especially for the Ti-O-O terminated
interface. With the bulk-like model, different terminations and
stacking sequences were also studied and the interface energy is
calculated with no surface energy involved. Except in the one case
with anomalous surface reconstruction, agreement between the two
models is good.

The relaxations tended to flatten the metallic and ionic surfaces.

Substantial charge transfer in the interface structure is observed,
giving interface charges exceeding those in bulk TiO$_2$ but this is confined to a small region around 5 \AA~ of the
interface.  As more electrons move to the oxide, the ionic bonding between the interfacial
Ti atoms in the metal and the interfacial O in the oxide becomes
stronger. These enhanced electrostatic fields at the interface are evidenced
by the shift of the low-lying 2s state in the pDOS. More electrons are accepted by the interfacial oxygen ions
than oxygens ions in the bulk rutile; this can be easily understood because the electron
donators are more abundant in the presence of the metal slab.

From the results of the projected DOS calculation, the picture emerges
of metallic bonding changing to ionic bonding over a short region,
with enhanced ionic bonding at the interface. In the O-O-Ti and O-Ti-O
interfaces, a small covalent component may exist, but for the Ti-O-O
interface, no hybridized states are observed on the interfacial Ti of
the metal slab, thus no covalent interfacial bonding is formed. Also
we notice that new gap states formed in the DOS of interfacial Ti of
the oxide. As with structure and Bader charge, the atom-projected DOS is consistent between the two models.

Most notably, the total energy of the slabs containing interfaces is lower
than the equivalent numbers of atoms in pure Ti and pure TiO$_2$.
This shows that Ti has strong affinity for oxygen, but at the same
time that the oxide is strongly bound to Ti.  Combined with the good
match of lattice parameters, this helps to explain why the oxide forms
such a good protective coating for Ti alloys.

\acknowledgement

The authors would like to acknowledge the support of the European Union Seventh Framework Programme
(FP7/2007-2013) under grant agreement No. PITN-GA-2008-211536, project MaMiNa. The authors also acknowledge the financial support from the MoST of China under Grant No. 2011CB606404.

\bibliography{interface}

\providecommand*\mcitethebibliography{\thebibliography}
\csname @ifundefined\endcsname{endmcitethebibliography}
  {\let\endmcitethebibliography\endthebibliography}{}
\begin{mcitethebibliography}{37}
\providecommand*\natexlab[1]{#1}
\providecommand*\mciteSetBstSublistMode[1]{}
\providecommand*\mciteSetBstMaxWidthForm[2]{}
\providecommand*\mciteBstWouldAddEndPuncttrue
  {\def\EndOfBibitem{\unskip.}}
\providecommand*\mciteBstWouldAddEndPunctfalse
  {\let\EndOfBibitem\relax}
\providecommand*\mciteSetBstMidEndSepPunct[3]{}
\providecommand*\mciteSetBstSublistLabelBeginEnd[3]{}
\providecommand*\EndOfBibitem{}
\mciteSetBstSublistMode{f}
\mciteSetBstMaxWidthForm{subitem}{(\alph{mcitesubitemcount})}
\mciteSetBstSublistLabelBeginEnd
  {\mcitemaxwidthsubitemform\space}
  {\relax}
  {\relax}

\bibitem[Lutjering and Williams(2007)Lutjering, and Williams]{Titanium}
Lutjering,~G., Williams,~J.~C., Eds. \emph{Titanium}, 2nd ed.; Springer:
  Germany, 2007\relax
\mciteBstWouldAddEndPuncttrue
\mciteSetBstMidEndSepPunct{\mcitedefaultmidpunct}
{\mcitedefaultendpunct}{\mcitedefaultseppunct}\relax
\EndOfBibitem
\bibitem[Flower and Swann(1974)Flower, and Swann]{Flower1974}
Flower,~H.; Swann,~P. \emph{Acta Metallurgica} \textbf{1974}, \emph{22}, 1339
  -- 1347\relax
\mciteBstWouldAddEndPuncttrue
\mciteSetBstMidEndSepPunct{\mcitedefaultmidpunct}
{\mcitedefaultendpunct}{\mcitedefaultseppunct}\relax
\EndOfBibitem
\bibitem[Kozlowski et~al.(1988)Kozlowski, Tyler, Smyrl, and
  Atanasoski]{Kozlowski1988}
Kozlowski,~M.; Tyler,~P.; Smyrl,~W.; Atanasoski,~R. \emph{Surface Science}
  \textbf{1988}, \emph{194}, 505 -- 530\relax
\mciteBstWouldAddEndPuncttrue
\mciteSetBstMidEndSepPunct{\mcitedefaultmidpunct}
{\mcitedefaultendpunct}{\mcitedefaultseppunct}\relax
\EndOfBibitem
\bibitem[Ting et~al.(2002)Ting, Chen, and Liu]{Ting2002}
Ting,~C.-C.; Chen,~S.-Y.; Liu,~D.-M. \emph{Thin Solid Films} \textbf{2002},
  \emph{402}, 290 -- 295\relax
\mciteBstWouldAddEndPuncttrue
\mciteSetBstMidEndSepPunct{\mcitedefaultmidpunct}
{\mcitedefaultendpunct}{\mcitedefaultseppunct}\relax
\EndOfBibitem
\bibitem[Kumar et~al.(2010)Kumar, Narayanan, Raman, and Seshadri]{Kumar2010}
Kumar,~S.; Narayanan,~T.~S.; Raman,~S. G.~S.; Seshadri,~S. \emph{Materials
  Characterization} \textbf{2010}, \emph{61}, 589 -- 597\relax
\mciteBstWouldAddEndPuncttrue
\mciteSetBstMidEndSepPunct{\mcitedefaultmidpunct}
{\mcitedefaultendpunct}{\mcitedefaultseppunct}\relax
\EndOfBibitem
\bibitem[Lopez et~al.(2003)Lopez, Jimenez, and Gutierrez]{Lopez2003}
Lopez,~M.; Jimenez,~J.; Gutierrez,~A. \emph{Electrochimica Acta} \textbf{2003},
  \emph{48}, 1395 -- 1401\relax
\mciteBstWouldAddEndPuncttrue
\mciteSetBstMidEndSepPunct{\mcitedefaultmidpunct}
{\mcitedefaultendpunct}{\mcitedefaultseppunct}\relax
\EndOfBibitem
\bibitem[Garcia-Alonso et~al.(2003)Garcia-Alonso, Saldana, Valles,
  Gonzalez-Carrasco, Gonzalez-Cabrero, Martinez, Gil-Garay, and
  Munuera]{Garcia2003}
Garcia-Alonso,~M.; Saldana,~L.; Valles,~G.; Gonzalez-Carrasco,~J.;
  Gonzalez-Cabrero,~J.; Martinez,~M.; Gil-Garay,~E.; Munuera,~L.
  \emph{Biomaterials} \textbf{2003}, \emph{24}, 19 -- 26\relax
\mciteBstWouldAddEndPuncttrue
\mciteSetBstMidEndSepPunct{\mcitedefaultmidpunct}
{\mcitedefaultendpunct}{\mcitedefaultseppunct}\relax
\EndOfBibitem
\bibitem[Guleryuz and Cimenoglu(2004)Guleryuz, and Cimenoglu]{Guleryuz2004}
Guleryuz,~H.; Cimenoglu,~H. \emph{Biomaterials} \textbf{2004}, \emph{25}, 3325
  -- 3333\relax
\mciteBstWouldAddEndPuncttrue
\mciteSetBstMidEndSepPunct{\mcitedefaultmidpunct}
{\mcitedefaultendpunct}{\mcitedefaultseppunct}\relax
\EndOfBibitem
\bibitem[Siegel et~al.(2002)Siegel, Hector, and Adams]{Siegel2002}
Siegel,~D.~J.; Hector,~L.~G.; Adams,~J.~B. \emph{Phys. Rev. B} \textbf{2002},
  \emph{65}, 085415\relax
\mciteBstWouldAddEndPuncttrue
\mciteSetBstMidEndSepPunct{\mcitedefaultmidpunct}
{\mcitedefaultendpunct}{\mcitedefaultseppunct}\relax
\EndOfBibitem
\bibitem[Christensen and Carter(2001)Christensen, and Carter]{Christ2001}
Christensen,~A.; Carter,~E.~A. \emph{The Journal of Chemical Physics}
  \textbf{2001}, \emph{114}, 5816--5831\relax
\mciteBstWouldAddEndPuncttrue
\mciteSetBstMidEndSepPunct{\mcitedefaultmidpunct}
{\mcitedefaultendpunct}{\mcitedefaultseppunct}\relax
\EndOfBibitem
\bibitem[Liu et~al.(2003)Liu, Wang, and Ye]{Liu2003}
Liu,~L.~M.; Wang,~S.~Q.; Ye,~H.~Q. \emph{Surface and Interface Analysis}
  \textbf{2003}, \emph{35}, 835--841\relax
\mciteBstWouldAddEndPuncttrue
\mciteSetBstMidEndSepPunct{\mcitedefaultmidpunct}
{\mcitedefaultendpunct}{\mcitedefaultseppunct}\relax
\EndOfBibitem
\bibitem[Lallet et~al.(2009)Lallet, Olivi-Tran, and Lewis]{Lallet2009}
Lallet,~F.; Olivi-Tran,~N.; Lewis,~L.~J. \emph{Phys. Rev. B} \textbf{2009},
  \emph{79}, 035413\relax
\mciteBstWouldAddEndPuncttrue
\mciteSetBstMidEndSepPunct{\mcitedefaultmidpunct}
{\mcitedefaultendpunct}{\mcitedefaultseppunct}\relax
\EndOfBibitem
\bibitem[Shang et~al.(2010)Shang, Guan, and Wang]{Shang2010}
Shang,~J.-X.; Guan,~K.; Wang,~F.-H. \emph{Journal of Physics: Condensed Matter}
  \textbf{2010}, \emph{22}, 085004\relax
\mciteBstWouldAddEndPuncttrue
\mciteSetBstMidEndSepPunct{\mcitedefaultmidpunct}
{\mcitedefaultendpunct}{\mcitedefaultseppunct}\relax
\EndOfBibitem
\bibitem[Kresse and Hafner(1993)Kresse, and Hafner]{Kresse1993}
Kresse,~G.; Hafner,~J. \emph{Phys. Rev. B} \textbf{1993}, \emph{47},
  558--561\relax
\mciteBstWouldAddEndPuncttrue
\mciteSetBstMidEndSepPunct{\mcitedefaultmidpunct}
{\mcitedefaultendpunct}{\mcitedefaultseppunct}\relax
\EndOfBibitem
\bibitem[Kresse and Furthm\"uller(1996)Kresse, and Furthm\"uller]{Kresse1996}
Kresse,~G.; Furthm\"uller,~J. \emph{Phys. Rev. B} \textbf{1996}, \emph{54},
  11169--11186\relax
\mciteBstWouldAddEndPuncttrue
\mciteSetBstMidEndSepPunct{\mcitedefaultmidpunct}
{\mcitedefaultendpunct}{\mcitedefaultseppunct}\relax
\EndOfBibitem
\bibitem[G. and J.(1996)G., and J.]{KresseG1996}
G.,~K.; J.,~F. \emph{Computational Materials Science} \textbf{1996}, \emph{6},
  15--50\relax
\mciteBstWouldAddEndPuncttrue
\mciteSetBstMidEndSepPunct{\mcitedefaultmidpunct}
{\mcitedefaultendpunct}{\mcitedefaultseppunct}\relax
\EndOfBibitem
\bibitem[Bl\"ochl(1994)]{Blochl1994}
Bl\"ochl,~P.~E. \emph{Phys. Rev. B} \textbf{1994}, \emph{50},
  17953--17979\relax
\mciteBstWouldAddEndPuncttrue
\mciteSetBstMidEndSepPunct{\mcitedefaultmidpunct}
{\mcitedefaultendpunct}{\mcitedefaultseppunct}\relax
\EndOfBibitem
\bibitem[Perdew and Wang(1992)Perdew, and Wang]{Perdew1992}
Perdew,~J.~P.; Wang,~Y. \emph{Phys. Rev. B} \textbf{1992}, \emph{46},
  12947--12954\relax
\mciteBstWouldAddEndPuncttrue
\mciteSetBstMidEndSepPunct{\mcitedefaultmidpunct}
{\mcitedefaultendpunct}{\mcitedefaultseppunct}\relax
\EndOfBibitem
\bibitem[Monkhorst and Pack(1976)Monkhorst, and Pack]{Monkhorst1976}
Monkhorst,~H.~J.; Pack,~J.~D. \emph{Phys. Rev. B} \textbf{1976}, \emph{13},
  5188--5192\relax
\mciteBstWouldAddEndPuncttrue
\mciteSetBstMidEndSepPunct{\mcitedefaultmidpunct}
{\mcitedefaultendpunct}{\mcitedefaultseppunct}\relax
\EndOfBibitem
\bibitem[Kittel(2004)]{Kittle1971}
Kittel,~C., Ed. \emph{Introduction to Solid State Physics}, 8th ed.; John Wiley
  and Sons: New York, 2004\relax
\mciteBstWouldAddEndPuncttrue
\mciteSetBstMidEndSepPunct{\mcitedefaultmidpunct}
{\mcitedefaultendpunct}{\mcitedefaultseppunct}\relax
\EndOfBibitem
\bibitem[Burdett et~al.(1987)Burdett, Hughbanks, Miller, Richardson, and
  Smith]{Burdett1987}
Burdett,~J.~K.; Hughbanks,~T.; Miller,~G.~J.; Richardson,~J.~W.; Smith,~J.~V.
  \emph{Journal of the American Chemical Society} \textbf{1987}, \emph{109},
  3639--3646\relax
\mciteBstWouldAddEndPuncttrue
\mciteSetBstMidEndSepPunct{\mcitedefaultmidpunct}
{\mcitedefaultendpunct}{\mcitedefaultseppunct}\relax
\EndOfBibitem
\bibitem[GRANT(1959)]{Grant1959}
GRANT,~F.~A. \emph{Rev. Mod. Phys.} \textbf{1959}, \emph{31}, 646--674\relax
\mciteBstWouldAddEndPuncttrue
\mciteSetBstMidEndSepPunct{\mcitedefaultmidpunct}
{\mcitedefaultendpunct}{\mcitedefaultseppunct}\relax
\EndOfBibitem
\bibitem[III and Watson(1989)III, and Watson]{Watson1989}
III,~J.~M.; Watson,~P. \emph{Surface Science Letters} \textbf{1989},
  \emph{209}, L105 -- L108\relax
\mciteBstWouldAddEndPuncttrue
\mciteSetBstMidEndSepPunct{\mcitedefaultmidpunct}
{\mcitedefaultendpunct}{\mcitedefaultseppunct}\relax
\EndOfBibitem
\bibitem[III and Watson(1989)III, and Watson]{Mischenko1989}
III,~J.~M.; Watson,~P. \emph{Surface Science} \textbf{1989}, \emph{220}, L667
  -- L670\relax
\mciteBstWouldAddEndPuncttrue
\mciteSetBstMidEndSepPunct{\mcitedefaultmidpunct}
{\mcitedefaultendpunct}{\mcitedefaultseppunct}\relax
\EndOfBibitem
\bibitem[Diebold(2003)]{Diebold2003}
Diebold,~U. \emph{Surface Science Reports} \textbf{2003}, \emph{48}, 53 --
  229\relax
\mciteBstWouldAddEndPuncttrue
\mciteSetBstMidEndSepPunct{\mcitedefaultmidpunct}
{\mcitedefaultendpunct}{\mcitedefaultseppunct}\relax
\EndOfBibitem
\bibitem[Boettger(1994)]{Boettger1994}
Boettger,~J.~C. \emph{Phys. Rev. B} \textbf{1994}, \emph{49},
  16798--16800\relax
\mciteBstWouldAddEndPuncttrue
\mciteSetBstMidEndSepPunct{\mcitedefaultmidpunct}
{\mcitedefaultendpunct}{\mcitedefaultseppunct}\relax
\EndOfBibitem
\bibitem[Erdin et~al.(2005)Erdin, Lin, and Halley]{Erdin2005}
Erdin,~S.; Lin,~Y.; Halley,~J.~W. \emph{Phys. Rev. B} \textbf{2005}, \emph{72},
  035405\relax
\mciteBstWouldAddEndPuncttrue
\mciteSetBstMidEndSepPunct{\mcitedefaultmidpunct}
{\mcitedefaultendpunct}{\mcitedefaultseppunct}\relax
\EndOfBibitem
\bibitem[Tyson and Miller(1977)Tyson, and Miller]{Tyson1977}
Tyson,~W.; Miller,~W. \emph{Surface Science} \textbf{1977}, \emph{62}, 267 --
  276\relax
\mciteBstWouldAddEndPuncttrue
\mciteSetBstMidEndSepPunct{\mcitedefaultmidpunct}
{\mcitedefaultendpunct}{\mcitedefaultseppunct}\relax
\EndOfBibitem
\bibitem[Muscat et~al.(1999)Muscat, Harrison, and Thornton]{Muscat1999}
Muscat,~J.; Harrison,~N.~M.; Thornton,~G. \emph{Phys. Rev. B} \textbf{1999},
  \emph{59}, 2320--2326\relax
\mciteBstWouldAddEndPuncttrue
\mciteSetBstMidEndSepPunct{\mcitedefaultmidpunct}
{\mcitedefaultendpunct}{\mcitedefaultseppunct}\relax
\EndOfBibitem
\bibitem[Labat et~al.(2008)Labat, Baranek, and Adamo]{Labat2008}
Labat,~F.; Baranek,~P.; Adamo,~C. \emph{Journal of Chemical Theory and
  Computation} \textbf{2008}, \emph{4}, 341--352\relax
\mciteBstWouldAddEndPuncttrue
\mciteSetBstMidEndSepPunct{\mcitedefaultmidpunct}
{\mcitedefaultendpunct}{\mcitedefaultseppunct}\relax
\EndOfBibitem
\bibitem[Perron et~al.(2007)Perron, Domain, Roques, Drot, Simoni, and
  Catalette]{Perron2007}
Perron,~H.; Domain,~C.; Roques,~J.; Drot,~R.; Simoni,~E.; Catalette,~H.
  \emph{Theoretical Chemistry Accounts: Theory, Computation, and Modeling
  (Theoretica Chimica Acta)} \textbf{2007}, \emph{117}, 565--574,
  10.1007/s00214-006-0189-y\relax
\mciteBstWouldAddEndPuncttrue
\mciteSetBstMidEndSepPunct{\mcitedefaultmidpunct}
{\mcitedefaultendpunct}{\mcitedefaultseppunct}\relax
\EndOfBibitem
\bibitem[Zhang and Smith(2000)Zhang, and Smith]{Zhang2000}
Zhang,~W.; Smith,~J.~R. \emph{Phys. Rev. Lett.} \textbf{2000}, \emph{85},
  3225--3228\relax
\mciteBstWouldAddEndPuncttrue
\mciteSetBstMidEndSepPunct{\mcitedefaultmidpunct}
{\mcitedefaultendpunct}{\mcitedefaultseppunct}\relax
\EndOfBibitem
\bibitem[ller et~al.(2010)ller, Kristoffersen, Hinnemann, Madsen, and
  rk~Hammer]{Jess2010}
ller,~J. S.-M.; Kristoffersen,~H.~H.; Hinnemann,~B.; Madsen,~G. K.~H.;
  rk~Hammer,~B. \emph{The Journal of Chemical Physics} \textbf{2010},
  \emph{133}, 144708\relax
\mciteBstWouldAddEndPuncttrue
\mciteSetBstMidEndSepPunct{\mcitedefaultmidpunct}
{\mcitedefaultendpunct}{\mcitedefaultseppunct}\relax
\EndOfBibitem
\bibitem[Tang et~al.(2009)Tang, Sanville, and Henkelman]{Tang2009}
Tang,~W.; Sanville,~E.; Henkelman,~G. \emph{Journal of Physics: Condensed
  Matter} \textbf{2009}, \emph{21}, 084204\relax
\mciteBstWouldAddEndPuncttrue
\mciteSetBstMidEndSepPunct{\mcitedefaultmidpunct}
{\mcitedefaultendpunct}{\mcitedefaultseppunct}\relax
\EndOfBibitem
\bibitem[Paxton and Thi\^en-Nga(1998)Paxton, and Thi\^en-Nga]{Paxton1998}
Paxton,~A.~T.; Thi\^en-Nga,~L. \emph{Phys. Rev. B} \textbf{1998}, \emph{57},
  1579--1584\relax
\mciteBstWouldAddEndPuncttrue
\mciteSetBstMidEndSepPunct{\mcitedefaultmidpunct}
{\mcitedefaultendpunct}{\mcitedefaultseppunct}\relax
\EndOfBibitem
\bibitem[Landa et~al.(2009)Landa, S\"oderlind, Ruban, Peil, and
  Vitos]{Landa2009}
Landa,~A.; S\"oderlind,~P.; Ruban,~A.~V.; Peil,~O.~E.; Vitos,~L. \emph{Phys.
  Rev. Lett.} \textbf{2009}, \emph{103}, 235501\relax
\mciteBstWouldAddEndPuncttrue
\mciteSetBstMidEndSepPunct{\mcitedefaultmidpunct}
{\mcitedefaultendpunct}{\mcitedefaultseppunct}\relax
\EndOfBibitem
\end{mcitethebibliography}

\begin{tocentry}
\includegraphics{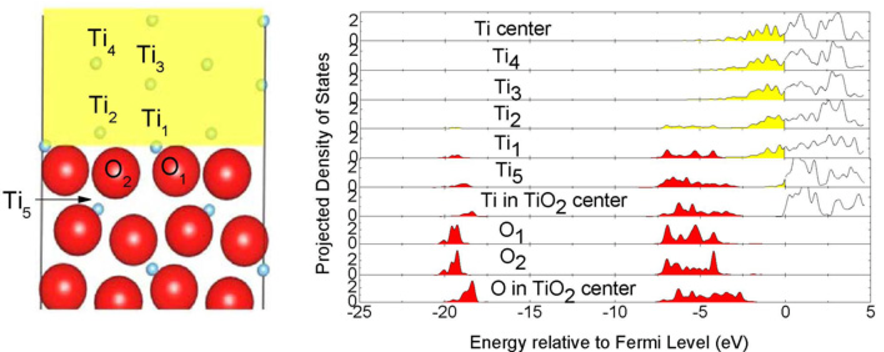}
\end{tocentry}

\end{document}